\documentclass[twocolumn,reprint,epsf,graphics,psfig,superscriptaddress]{revtex4}
\usepackage[usenames,dvipsnames]{color}
\usepackage{amsfonts}
\usepackage{graphicx}
\usepackage{amsmath}
\usepackage{amssymb}
\usepackage{latexsym}
\usepackage{psfrag}
\usepackage{dcolumn}
\usepackage{datetime}
\usepackage{multirow}
\usepackage{subfigure}
\usepackage{soul}

\begin{document}
\title{Coulomb bound states and resonances due to groups of Ca dimers adsorbed
on suspended graphene}
\author{Alireza Saffarzadeh}
\altaffiliation{Author to whom correspondence should be addressed. Electronic mail: asaffarz@sfu.ca}
\affiliation{Department of Physics, Payame Noor University, P.O.
Box 19395-3697 Tehran, Iran} \affiliation{Department of Physics,
Simon Fraser University, Burnaby, British Columbia, Canada V5A
1S6}
\author{George Kirczenow}
\affiliation{Department of Physics, Simon Fraser University,
Burnaby, British Columbia, Canada V5A 1S6}
\date{\today}

\begin{abstract}
The electronic bound states and resonances in the vicinity of the
Dirac point energy due to the adsorption of calcium dimers on a
suspended graphene monolayer are explored theoretically using
density functional theory (DFT) and an improved extended
H\"{u}ckel model that includes electrostatic potentials. The
Mulliken atomic charges and the electrostatic potentials are
obtained from DFT calculations and reveal charge transfer from the
Ca dimers to the graphene which is responsible for the emergence
of resonant states in the electronic spectrum. The number of
resonant states increases as the number of adsorbed dimers is
increased. We find a bound ``atomic-collapse" state in the
graphene local density of states, as has been observed
experimentally [Wang \textit{et al.}, Science {\bf 340}, 734
(2013)]. We find the formation of the atomic-collapse state and
its population with electrons to require fewer adsorbed Ca dimers
than in the experiment, possibly due to the different spacing
between dimers and the dielectric screening by a boron nitride
substrate in the experiment. We also predict the onset of filling
of a second atomic-collapse state with electrons when six Ca
dimers are adsorbed on the suspended graphene monolayer.
Experiments testing these predictions would be of interest.
 \end{abstract}
\maketitle

\section{Introduction}
The electronic properties of graphene have attracted a great deal
of research due to their fundamental interest and promising
applications \cite{NZGG} which are related to graphene's linear
dispersion near the Dirac point energy \cite{Wallace,Neto}. In
this regard, the problem of charged impurities in graphene has
recently received considerable attention
\cite{Pereira,Shytov1,Shytov2,Fogler,Wang1}. These impurities, as
well as adsorbates, can induce resonant states in the vicinity of
the Dirac point energy that play an important role in a number of
properties of graphene nanostructures, such as transport
\cite{Ando,Nomura,Hwang1,Ihnatsenka,review}, chemical sensing
\cite{Schedin,Hwang2,Wehling}, controlled doping \cite{Ohta,Uchoa}
and magnetism \cite{Mao,Chan0,Wehling5,Wehling6,
Can,Jacob,Chan,Rappoport,Wehling8,Liu,Saffarzadeh}.

It was predicted theoretically that when the Coulomb potential
strength for heavily charged impurities in graphene exceeds a
certain critical value, an infinite family of Rydberg-like
quasi-bound states appears abruptly \cite{Shytov1}. Therefore, as
an atomic collapse (fall to the center effect) was predicted for
isolated atoms with highly charged nuclei, one can expect a
similar situation with highly charged impurities in graphene
\cite{Shytov1}. Such a phenomenon was observed recently in the
scanning tunneling microscopy (STM) experiments of Wang \textit{et
al.} \cite{Wang} in ultra-high vacuum. Using atomic manipulation,
they assembled groups of dimers of calcium ions on the surface of
graphene placed on a boron nitride (BN) substrate. Ca dimers were
added to the group, one by one, until the total charge transferred
from the Ca to the graphene surpassed a critical threshold
\cite{Wang}. The measured spatial and energetic characteristics of
the electronic states around the Ca-dimer cluster showed the
emergence of an oscillation in the local density of states (LDOS)
that manifested itself as a resonance above the graphene Dirac
point as the number of dimers in the cluster was increased
\cite{Wang}. When the resonance shifted below the Dirac point, the
quasi-bound state was interpreted as the atomic-collapse
eigenstate \cite{Wang}. In support of this interpretation of the
experimental results, calculations were performed  \cite{Wang} to
simulate the LDOS of graphene near charged impurities by means of
a parameterized model for charged impurities and the
two-dimensional (2D) continuum Dirac model
\cite{Pereira,Shytov1,Shytov2}. An \textit{ab initio} density
functional theory (DFT) calculation was performed \cite{Wang} for
a {\em single} Ca dimer adsorbed on graphene to determine a
reasonable value for the model fitting parameter. These model
calculations \cite{Wang} were able to reproduce the main effect
observed in the experiment \cite{Wang}, namely, formation of the
the bound electronic state that is the analog of the
atomic-collapse phenomenon. However, a more detailed theoretical
treatment of the graphene electronic states associated with groups
of Ca dimers adsorbed on the graphene is clearly desirable.
Specifically, \textit{ab initio} DFT calculations of the charge
transfer and electrostatic potentials associated with more than
just one adsorbed Ca dimer (the case treated with DFT in Ref.
\onlinecite{Wang}), and calculations of the resonant and bound
electronic states based on atomistic graphene models, as distinct
from the continuum model considered in Ref. \onlinecite{Wang},
have not been available. Furthermore, the previous theoretical
study \cite{Wang} did not address the dependence of the spatial
and energetic characteristics of the resonant states on the
spatial arrangement of the Ca dimers on graphene.

In this paper, we explore the resonance features seen in the STM
experiment \cite{Wang} by means of DFT calculations of the relaxed
geometries, charge transfer between the Ca dimers and graphene,
and the associated electrostatic potentials for groups of 1-6 Ca
dimers adsorbed on suspended graphene. In order to model the
electronic structures for groups of Ca dimers adsorbed on graphene
substrates with {\em large} numbers of carbon atoms atomistically,
we use the (tight-binding) extended H\"{u}ckel model of quantum
chemistry \cite {George,Ammeter,Yaehmop} augmented by the
inclusion of electrostatic potentials obtained from DFT
calculations. The LDOS for graphene in the presence of 1-6 Ca
dimers is calculated and the emergence and evolution of resonances
and their spatial characteristics with increasing the numbers of
Ca dimers are discussed. In addition, we compare the Dirac point
resonances for six-dimer clusters (not addressed in the STM
experiment of Wang \textit{et al.} \cite{Wang}) with two different
dimer arrangements. Our calculations show that the resonance
features in the LDOS are qualitatively consistent with the
experiment and that the emergence of these resonances is due to
the charge transfer from Ca atoms to graphene monolayer and not
due to the atomic orbitals localized on the Ca atoms. Furthermore,
we find that the arrangement of the Ca dimers on the graphene can
affect the energies and oscillations of the resonances somewhat at
short distances from the centers of the Ca-dimer clusters.

However, we predict formation of the atomic-collapse state and its
population with electrons to require fewer Ca dimers in the
cluster adsorbed on the graphene than was required in the
experiment of Wang \textit{et al.} \cite{Wang}. We attribute this
difference to dielectric screening by the BN substrate underlying
the graphene in the experiment and to the larger spacing between
the Ca dimers in the clusters studied in the experiment.
\cite{Wang} Furthermore, for six adsorbed dimers (a case not
addressed in Ref. \onlinecite{Wang}), we predict a second
atomic-collapse state to also become partly populated with
electrons. Experimental tests of these predictions for compact
clusters of Ca dimers on suspended graphene would be of interest.

The paper is organized as follows. In Sec. II, we present the
model Hamiltonian for Ca dimers adsorbed on graphene, derived from
extended H\"{u}ckel theory augmented by the inclusion of
electrostatic potentials obtained from DFT calculations. We also
discuss the procedure for designing large Ca dimer/graphene
clusters for the purpose of eliminating edge effects in the
electronic states close to the graphene Dirac point. The Mulliken
atomic charges and the electrostatic potentials on the graphene
monolayer, obtained from DFT calculations in the presence of Ca
dimers, are discussed in Sec. III.A. The calculated LDOS within an
energy window close to the Dirac point of graphene at different
distances from the centers of the Ca-dimer clusters and
discussions of the resonant states, the atomic-collapse state, and
state filling with electrons are presented in Sec. III.B. In Sec.
III.C, the dependence of resonance states and their filling with
electrons on the spatial arrangement of the dimers is examined by
considering a six-dimer cluster and a six-dimer ring on the
graphene monolayer. Finally, in Sec. IV, we conclude with a
general discussion of our results.
\begin{figure}[t!]
\includegraphics[width=0.88\linewidth]{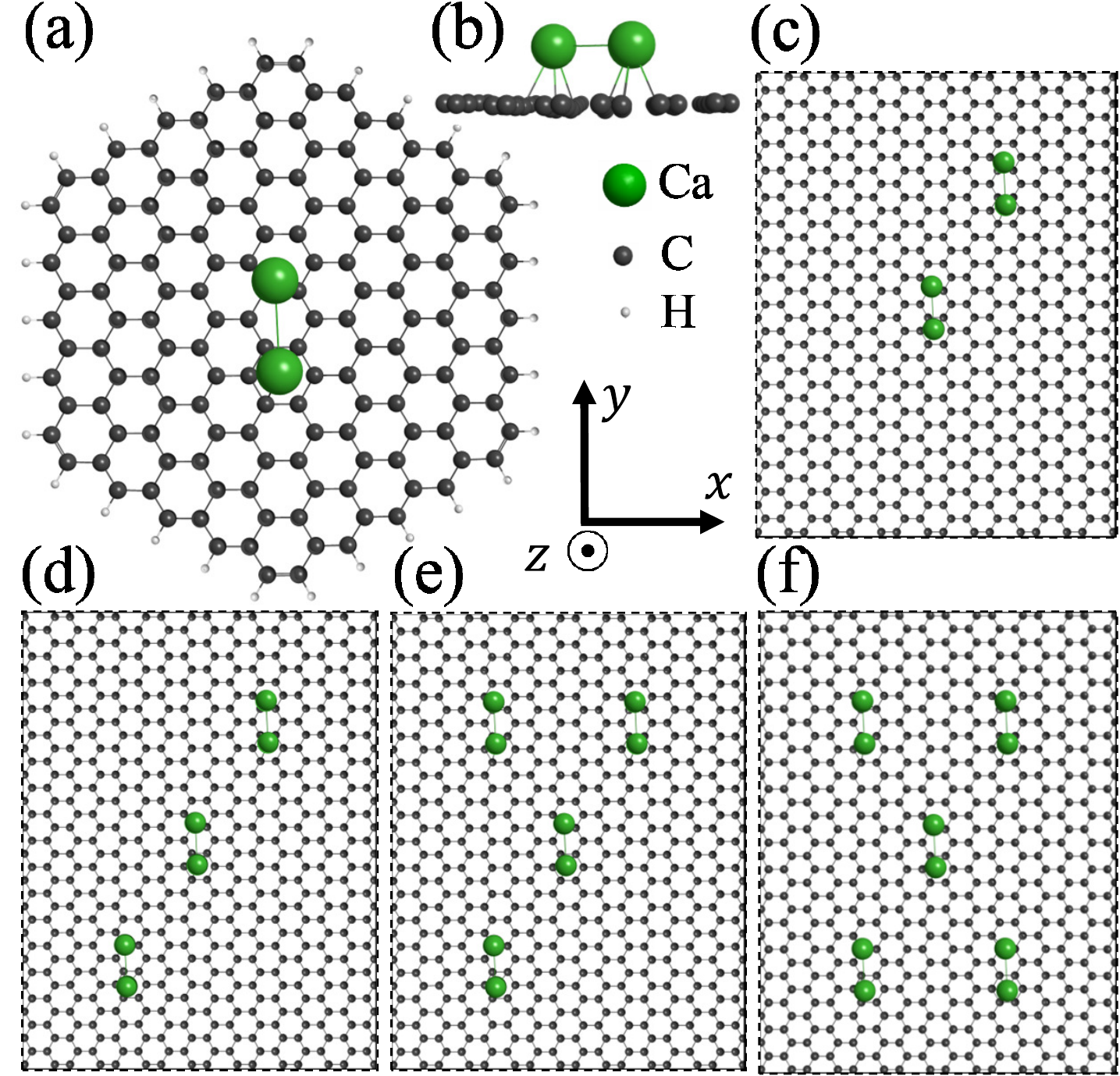}
\caption{(Color online) (a) The optimized geometrical structure of
a single Ca dimer on a graphene disk of 150 carbon atoms
passivated by 30 hydrogen atoms. (b) Side view of (a), but only
with 54 C atoms which shows the ($<$ 0.31 \AA) out-of-plane
(downward) distortion for the carbon atoms beneath the Ca dimer.
For the sake of clarity, the C-C bonds are not shown here. (c)
Two, (d) three, (e) four, and (f) five Ca dimers on a graphene
ribbon extended along the $y$-direction and designed based on the
optimized geometry shown in (a), and sandwiched between two
semi-infinite leads extended along the $x$-direction. Only a
segment of each central ribbon, referred to as the
``dimer-graphene cluster," is shown here.}\label{Figure1}
\end{figure}

\section{Theory}
We carried out \textit{ab initio} geometry relaxations based on
DFT for a single calcium dimer on a honeycomb graphene lattice
using the GAUSSIAN 09 software package with the PBEh1PBE hybrid
function and the 6-31G** basis set \cite{PBE}. For this purpose we
adopted a graphene disk consisting of 150 carbon atoms passivated
at the edges with 30 hydrogen atoms, the Ca dimer being bonded to
the graphene disk at its center. Although a finite-size graphene
cluster was used for the optimization calculations, the bond
lengths between neighboring C atoms were well converged with
increasing distance from the Ca,  which indicates that the number
of 150 C atoms is large enough to model the graphene monolayer.
Since all of the atoms were allowed to relax freely, we found a
maximum in-plane distortion of 0.02 {\AA} between carbon atoms and
a maximum out-of-plane distortion of 0.31 {\AA} (away from the Ca)
due to the presence of Ca dimer on the graphene. The final
optimized geometry obtained in this study is shown in Fig. 1(a).
The distance between the two Ca atoms of the dimer is 3.80 {\AA}
and the center of the dimer is located at 2.4 {\AA} above the
plane defined by the C atoms far from the Ca which is in agreement
with the recent \textit{ab initio} DFT calculations in a
generalized gradient approximation for a Ca dimer adsorbed on
graphene in supercells with periodic boundary conditions
\cite{Wang}. The positions of the Ca atoms in the graphene plane
(relative to the center of the graphene disk) are
$\mathbf{R}_1=-0.38\mathrm{\AA}\hat{x}+1.58\mathrm{\AA}\hat{y}$,
and
$\mathbf{R}_2=-0.20\mathrm{\AA}\hat{x}-2.21\mathrm{\AA}\hat{y}$,
indicating that the Ca atoms are in nonequivalent positions.

Since the adsorbed Ca dimers on graphene behave as charged
impurities, the electrostatic properties such as spatial charge
distributions and electrostatic potentials at each atomic site
should be computed by means of DFT calculations. For this reason,
we designed a larger graphene hexagonal disk similar to that shown
in Fig. 1(a) but with 1370 C atoms (not shown here). Then, we
transferred the atomic coordinates of the relaxed Ca dimer along
with those of the 54 closest C atoms surrounding the dimer and
belonging to the optimized cluster in Fig. 1(a) to the locations
of each of the Ca dimers of the larger disk to produce the
structures with multiple Ca dimers shown in Fig.1(c)-(f). This
procedure is justified since the distortion of the graphene
lattice induced by an adsorbed Ca dimer has a shorter range than
the distances between the Ca dimers in the clusters considered in
this work (see Fig. 1(b)).The cluster geometries formed in this
way were used to find the Mulliken atomic charges and the electric
potentials on the carbon atoms of the graphene and the calcium
atoms by means of DFT calculations.

The main difficulty in recognizing the physical resonant states
induced by Ca dimers, especially at large distances from the dimer
region where the influence of resonant states is not strong, is
the mixing of these states with fluctuations in the electronic
states around Dirac point energy due to edge effects in the model
system. In order to suppress these quantum size effects as much as
possible, a large rectangular graphene sheet (called here the
``central ribbon") consisting of 3940 carbon atoms with length
41.2 {\AA} in the $x$-direction and width 240 {\AA} in the
$y$-direction was designed. Then all the relevant optimized
coordinates and the atomic electrostatic potentials obtained from
the DFT calculations for the 1370 C atom disk with adsorbed Ca
dimers were applied to the appropriate atoms in this ribbon. Next,
we attached two semi-infinite graphene ribbons (``leads''
hereafter) with width 240 {\AA} to the left and right sides of the
central ribbon to produce an infinite ribbon (in the
$x$-direction) with width 240 {\AA}. It is assumed that, the
electrostatic potential values for all carbon atoms located far
away from the dimers region converge to a fixed value. Therefore,
for the remaining carbon atoms in the central ribbon and for all
the atoms in the semi-infinite leads a cut-off in the
electrostatic potential was used. A discussion of this point will
be given in the next section.

According to the description given above, the Hamiltonian of the
system which is partitioned in three blocks, i.e. the central
ribbon $C$ and the two semi-infinite leads at its left $(L)$ and
right $(R)$ sides, can be written as
\begin{equation}\label{1}
H=H_C+H_L+H_R+V_{LC}+V_{RC}\ ,
\end{equation}
where the Hamiltonians of the central ribbon, $H_C$, and the
leads, $H_{L,R}$, are described by a tight-binding model derived
from the extended H\"{u}ckel theory in a basis of C $2s$ and $2p$,
and Ca $4s$ and $4p$ valence orbitals, that can be written as
\begin{equation}\label{2}
H_{i\in\{C,L,R\}}=\sum_{\alpha}\epsilon_{i\alpha} d_{i\alpha}^\dag
d_{i\alpha}+ \sum_{\alpha,\beta}\gamma_{i,\alpha
\beta}(d_{i\alpha}^\dag d_{i\beta}+\mathrm{H.c.})\ ,
\end{equation}
where $d_{i\alpha}^\dag(d_{i\alpha})$ is the creation
(annihilation) operator for an electron in the $\alpha$th atomic
valence orbital $\psi_{i\alpha}$. Within extended H\"{u}ckel
theory $\epsilon_{i\alpha}$ is the experimentally determined
valence orbital ionization energy, and $\gamma_{i,\alpha\beta}$ is
the off-diagonal matrix element between valence orbitals
$\psi_{i\alpha}$ and $\psi_{i\beta}$. In Eq. (\ref{1}),
$V_{L(R)C}$ describes the coupling matrix between the lead $L$
$(R)$ and the central ribbon $C$. Since the basis set used in
extended H\"{u}ckel theory is nonorthogonal, the orbital overlap
$S_{i,\alpha \beta}=\langle\psi_{i\alpha}|\psi_{i\beta}\rangle$
can be non-zero and will be included in this study
\cite{Emberly1,Emberly2}. To include the atomic electrostatic
potentials obtained from our DFT calculations in the present
model, the diagonal and off-diagonal matrix elements in the
Hamiltonians are modified as
$\epsilon_{i\alpha}\rightarrow\epsilon_{i\alpha}+U_{i}(x_\alpha,y_\alpha)$
and
$\gamma_{i,\alpha\beta}\rightarrow\gamma_{i,\alpha\beta}+\frac{1}{2}[U_{i}(x_\alpha,y_\alpha)+U_{i}(x_\beta,y_\beta)]S_{i,\alpha\beta}$
where $x_\alpha(y_\alpha)$ is the $x$ ($y$) component of the
position of atom with orbital $\alpha$, and $U(x_\alpha,y_\alpha)$
is the electric potential energy (EPE) for an atom at point
$R\equiv(x_\alpha,y_\alpha)$. Note that the EPE values that will
be discussed below, are the same for all valence orbitals
localized on each given atom of the dimer-graphene system.

The Green's function of the central ribbon coupled to the graphene
leads can be written as \cite{Datta,Saffar2010}
\begin{equation}\label{3}
G_{C}(\epsilon)=[zS_C-H_C-\Sigma_L(\epsilon)-\Sigma_R(\epsilon)]^{-1}\
,
\end{equation}
where, $z=\epsilon+i\eta$ with $\eta$ being a positive
infinitesimal number, and $\Sigma_{\{L,R\}}$ are the self-energy
terms due to the semi-infinite leads, each of which is considered
as a semi-infinite stack of principal layers (ribbons) with
nearest-neighbor interactions \cite{Lopez} and can be defined as
\cite{Nardelli12}
\begin{equation}\label{4}
\Sigma_{L(R)}(\epsilon)=H_{L(R)C}^{\dag}g_{L(R)}(\epsilon)H_{L(R)C}\
,
\end{equation}
where $H_{L(R)C}=V_{L(R)C}-\epsilon S_{L(R)C}$ with $S_{L(R)C}$
being the overlap matrix between lead $L(R)$ and the central
ribbon $C$. $g_{L(R)}=(zS_{L(R)}-H_{L(R)})^{-1}$ represents the
surface Green's function of the lead $L(R)$ that can be written in
terms of appropriate transfer matrices and computed using the
iterative procedure introduced by L\'{o}pez-Sancho \textit{et al.}
\cite{Lopez,Nardelli12,Saffar2010}. Accordingly, we can calculate
the local density of states (LDOS) for the $n$th site in the
central ribbon from the diagonal elements of $G_C$ as
\begin{equation}\label{6}
\rho_n(\epsilon)=-\frac{1}{\pi}\mathrm{Im}[G_C(\epsilon)S_C]_{n,n}\
.
\end{equation}
Note that in all LDOS graphs that will be presented in this work,
the average of LDOS associated with two neighboring carbon atoms
in a single graphene unit cell is taken.

\begin{figure}[t!]
\includegraphics[width=0.65\linewidth]{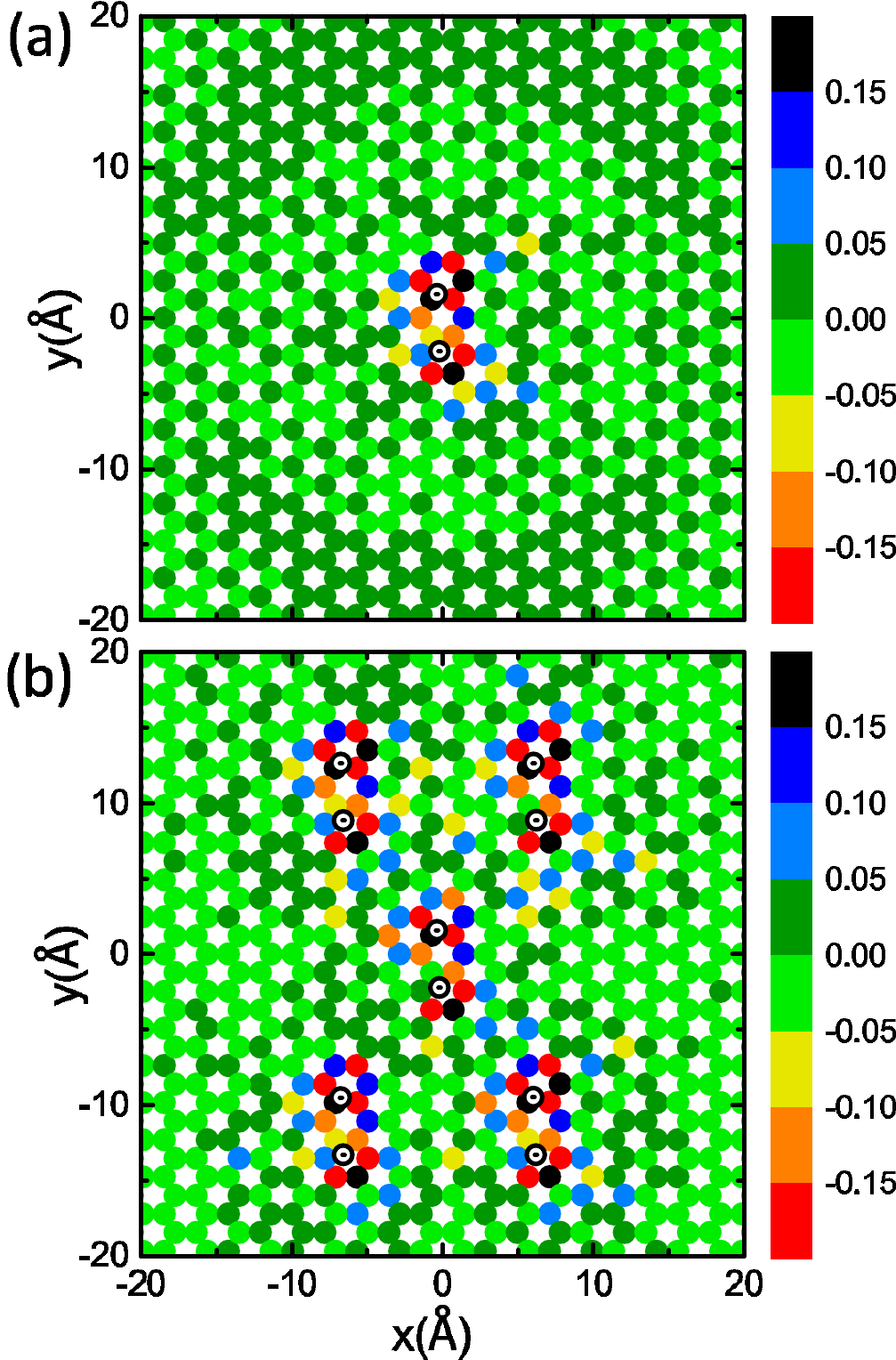}
\caption{(Color online) Distribution of Mulliken atomic charges
(in units of $|e|$) obtained from DFT for (a) one dimer and (b)
five dimers on graphene monolayer. The symbol $\odot$ shows the
location of Ca atoms in each dimer. The Ca atoms are positively
charged after charge transfer from the dimers to a graphene
monolayer.} \label{Figure2}
\end{figure}

\begin{figure}[t!]
\includegraphics[width=0.95\linewidth]{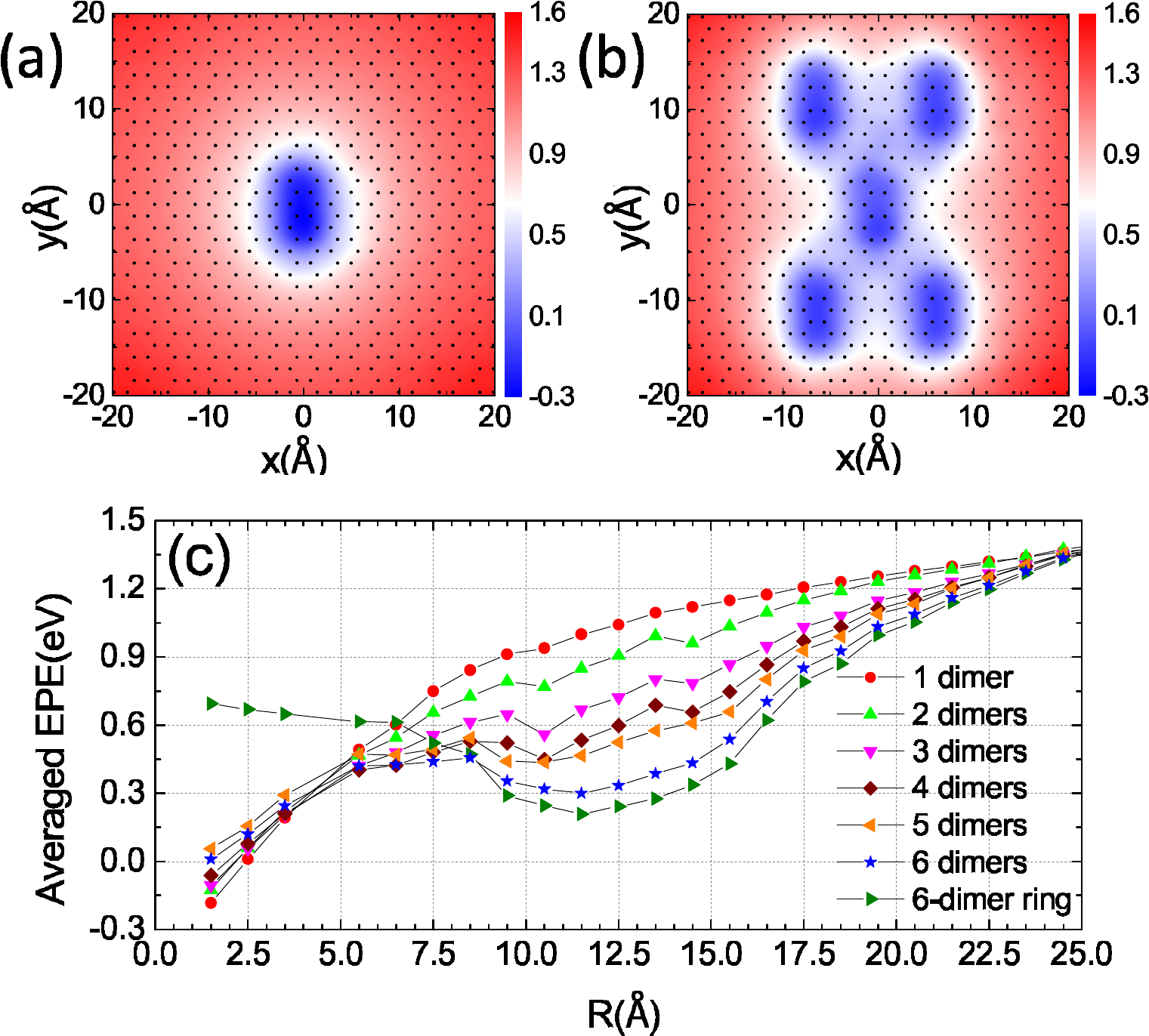}
\caption{(Color online) Distribution of atomic EPE for (a) one
dimer and (b) five dimers on a graphene monolayer. The black dots
in (a) and (b) show the locations of the carbon atom nuclei. (c)
The averaged EPE, $\bar{U}(R)$, as a function of distance from the
center of dimer-graphene clusters composed of 1-6 Ca dimers. Note
that there are two clusters with different spatial arrangements of
six Ca dimers, as discussed in Sec. IIIC.} \label{Figure3}
\end{figure}

\section{Results and discussion}
\subsection{Mulliken charges and electrostatic potentials}
Adsorption of Ca dimers on the graphene plane results in a charge
transfer between the Ca atoms of the dimers and the C atoms of the
graphene. To explore the nature of this charge transfer and the
polarity of charges, the Mulliken atomic charges were computed by
DFT for all of the dimer-graphene clusters. The results for the
case of one and five dimers are shown in Figs. 2(a) and 2(b),
respectively. Figure 2 shows that electronic charge is transferred
from the Ca dimers into graphene states and the calcium atoms
become positively charged, in agreement with the experiment
\cite{Wang}. In addition, the negative localized charges, shown by
light green circles around the single dimer in Fig. 2(a), form a
stadium-shaped region similar to the charge density seen in the
supplementary materials in Ref. \cite{Wang}. From the charge
distributions, it is obvious that the Mulliken atomic charges of
the carbon atoms in the neighborhood of each single calcium atom
are typically more negative than those of the carbon atoms far
away from the dimers, screening the bare charges on calcium atoms.
Some evidence of Friedel-like charge oscillations in the graphene
is also visible, especially in Fig. 2(a). Importantly, increasing
the number of dimers on the graphene sheet does not appreciably
affect the negative-charge accumulations on the carbon atoms
located in close vicinity to the Ca atoms, while in the case of a
five-dimer structure, shown in Fig. 2(b), the concentration of
negative charges on the carbon atoms away from the dimers is
higher than for the one-dimer case in Fig. 2(a). The average
charge transferred per Ca atom to the graphene depends on the
number of dimers adsorbed on the graphene. From the calculated
values of the Mulliken charges we found that as the number of
dimers in a cluster was increased from one to five, the average
charge transferred to the graphene per Ca atom in the cluster
changes as -0.83$|e|$, -0.72$|e|$, -0.70$|e|$, -0.69$|e|$, and
-0.69$|e|$, respectively. That is, somewhat less than one electron
per Ca atom is transferred to the graphene and the amount of
charge transferred per Ca atom decreases as the number of Ca atoms
in the cluster increases, as may be expected due to the repulsive
nature of the interaction between the electrons transferred to the
graphene.

The spatial distributions of the EPE, $U(x,y)=-|e|\Delta V$, for
two typical Ca-dimer clusters composed of one and five dimers are
shown in Figs. 3(a) and 3(b). Here, $e$ is the charge of electron
and $\Delta V=V_{c}(x,y)-V_{0}$ is the difference between electric
potential $V_{c}$ for an atom at point $R$ in the dimer-graphene
cluster containing 1370 carbon atoms, and $V_{0}$ due to an atom
of the same type as that at point $R$ in the cluster but in free
space. $R$ is measured with respect to the center of each of the
dimer-graphene clusters shown in Fig. 1. All of the electric
potentials obtained from DFT were computed with the GAUSSIAN 09
package. Note that only the EPE at the sites of carbon atoms of
the graphene are shown in Fig. 3. The EPE at the location of
carbon atoms clearly reveals the influence of adsorbed dimers on
the graphene in both one- and five-dimer clusters. The atomic
sites with negative (positive) EPE values represent positive
(negative) charges, which reflects the charge transfer from Ca
dimers to carbon atoms of the graphene, as expected. For $R\geq
15${\AA} away from the dimers region in the $x$ direction, the EPE
shows a convergence behavior due to the presence of screened
charges on the Ca dimers. In order to get a better understanding
of the behavior of EPE as a function of
$R=(x^2+y^2)^{\frac{1}{2}}$, we computed the average of EPE within
each $\Delta R=1${\AA} annulus for the dimer-graphene systems
composed of 1-6 Ca dimers (a discussion for a six-dimer cluster
and a six-dimer ring will be given in Section IIIC). The results
are shown in Fig. 3(c). The averaged EPE, $\bar{U}(R)$, for a
single Ca dimer on graphene rises monotonically with distance $R$
from the cluster center, as expected for the screened Coulomb
potential due to a positively charged dimer. As the number of Ca
dimers increases from one to five the electric potential in the
interval 5{\AA}$<R<$25{\AA} becomes deeper due to the charge
increase in the region of the dimers, which induces more
quasi-bound states and stronger resonances in the electronic
spectra of the dimer-graphene system. It is clear that the
asymptotic behavior of $\bar{U}(R)$ is almost linear for small and
large distances within the range shown in Fig. 3(c). For example,
in the case of a cluster with a single Ca dimer, the expressions
$\bar{U}(R)=0.18R-0.45$ and $\bar{U}(R)=0.025R+0.76$ ($\bar{U}$ is
in units of eV and $R$ is in \AA) can be fitted to the numerical
results in the range of $R\leq 5$\AA and 15\AA$\leq$R$\leq$25\AA
respectively.
\begin{figure}[t!]
\includegraphics[width=0.7\linewidth]{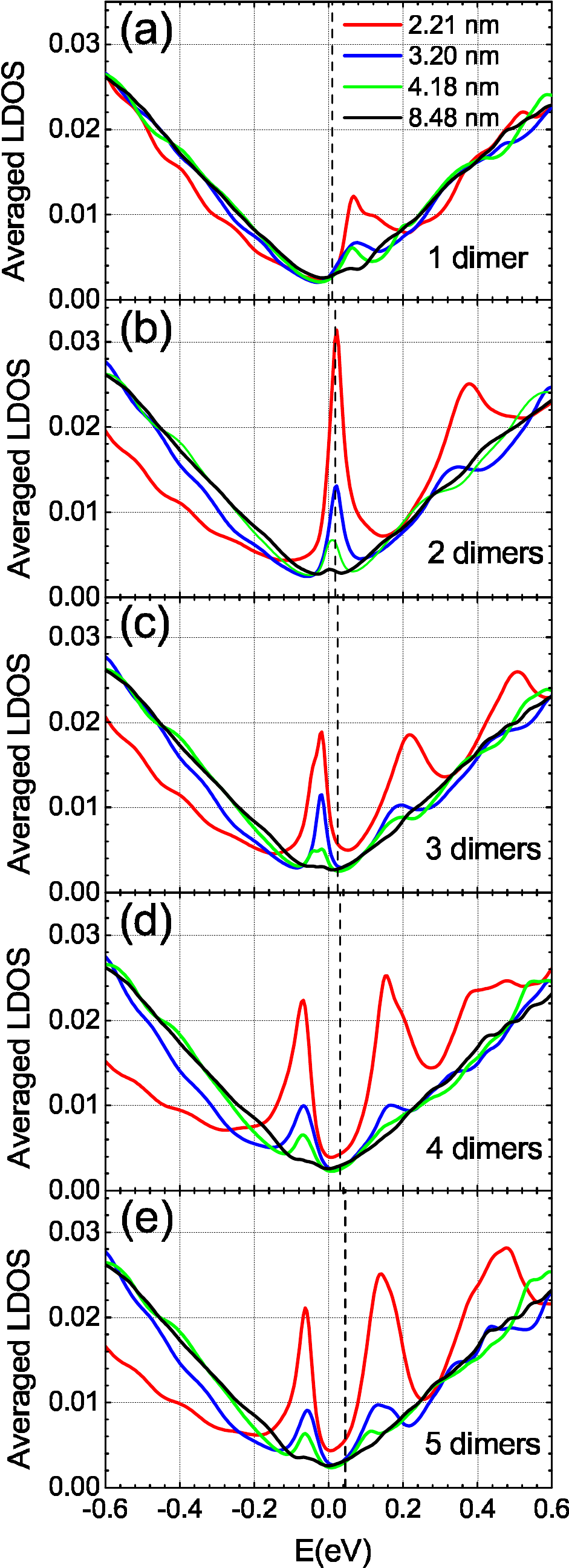}
\caption{(Color online) Calculated averaged LDOS for graphene in
the presence of 1-5 Ca dimers at different distances along the
$y$-axis and away from center of clusters shown in Fig.
\ref{Figure1}. The black dashed lines show the Fermi
energy.}\label{Figure4}
\end{figure}

The finite size of our 1370-carbon atom cluster used in the
electrostatic potential calculations results in potential
fluctuations at the edges of the cluster due to the presence of
electronic states located at the edges. Accordingly, in the
calculations reported below, we introduce a cutoff in the
potential profile to lessen the influence of such edge effects as
much as possible. The EPE cutoff value, chosen in all calculations
that follow, is 1.2 eV.

\begin{figure}[t!]
\includegraphics[width=0.8\linewidth]{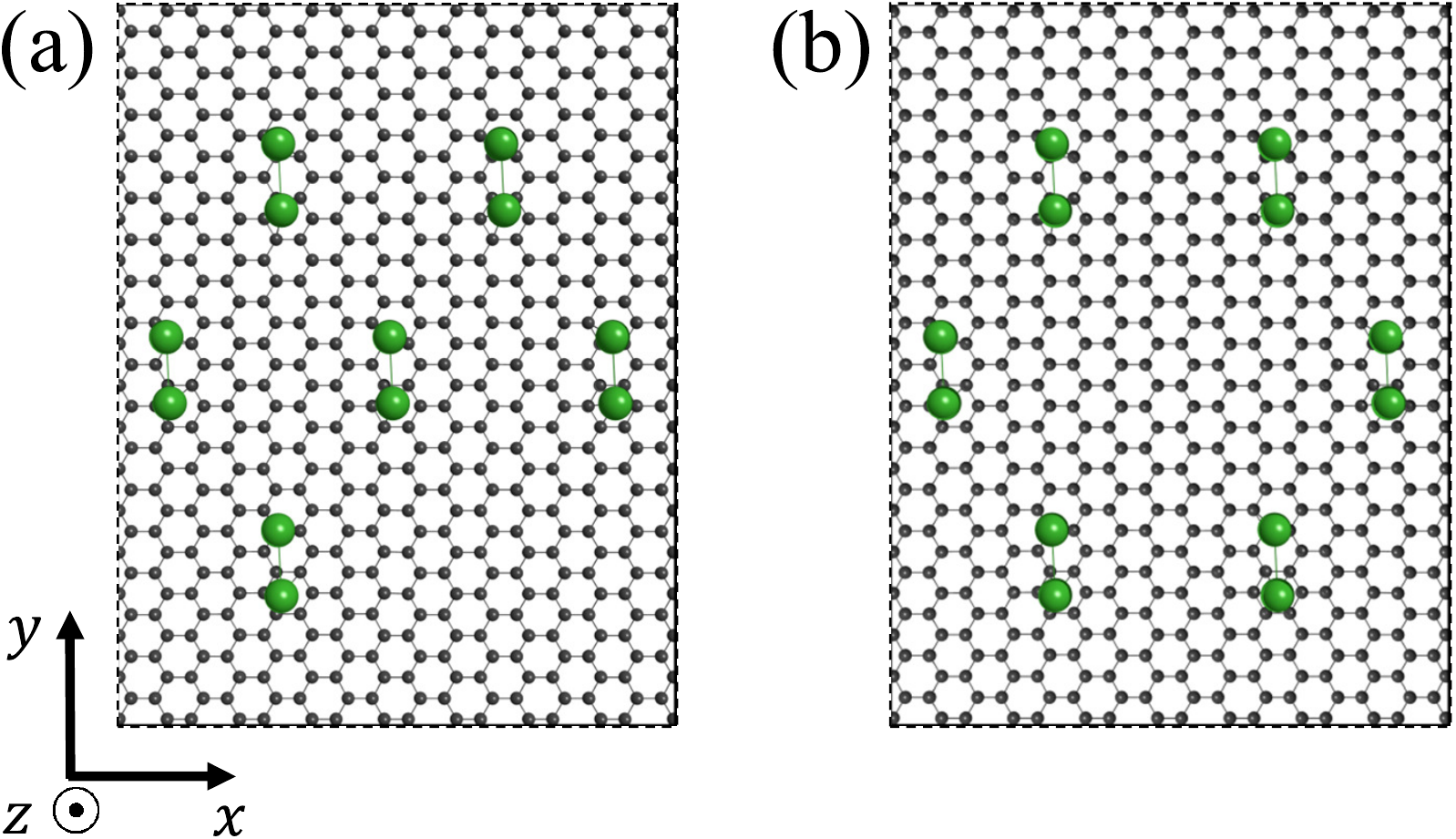}
\caption{(Color online) Two different arrangements of six Ca
dimers on a graphene plane. In the text, they are referred to as
(a) the six-dimer cluster and (b) the six-dimer ring, for
simplicity. Note that the center-center distance between two
adjacent dimers is 12.78 \AA.}\label{Figure6}
\end{figure}

\begin{figure}[t!]
\includegraphics[width=0.7\linewidth]{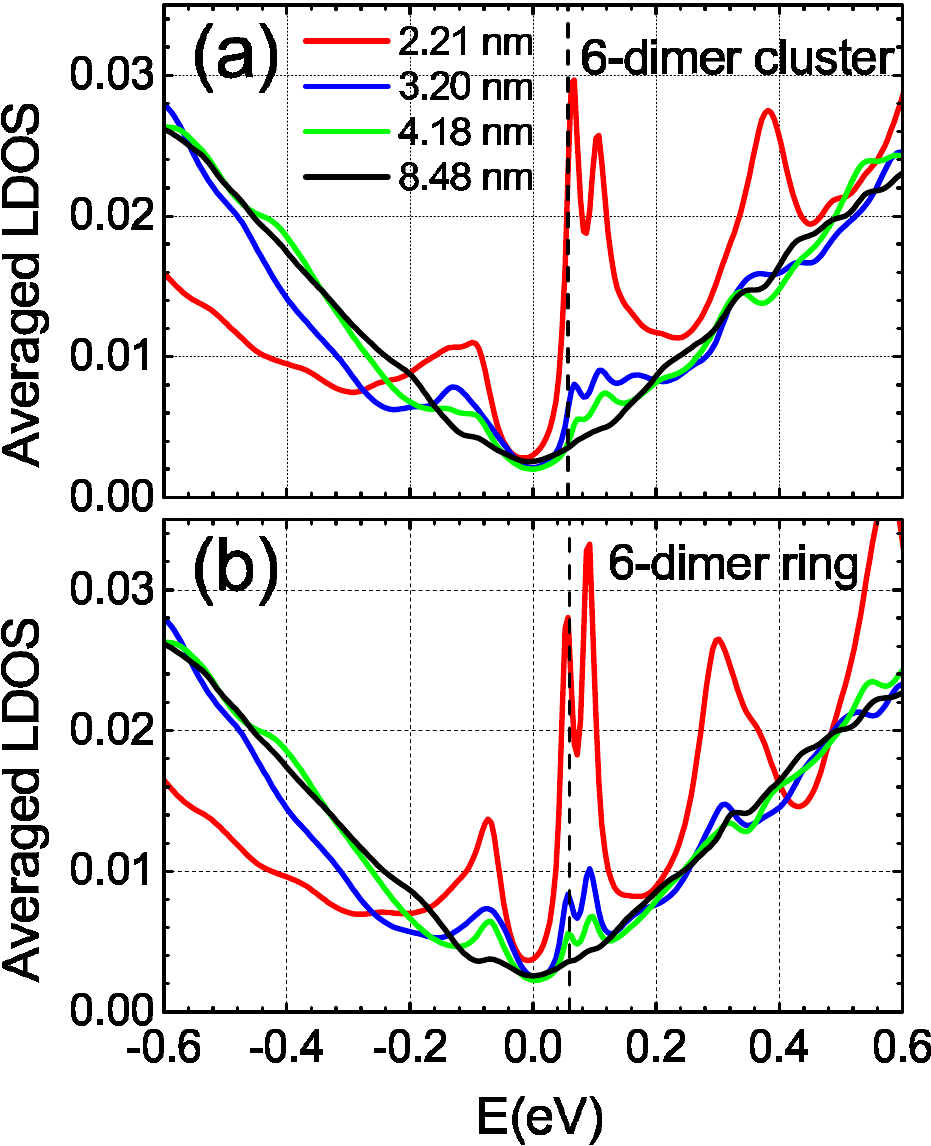}
\caption{(Color online) Calculated graphene LDOS in the presence
of (a) a six-dimer cluster and (b) a six-dimer ring at different
distances from the centers of the graphene clusters shown in Fig.
\ref{Figure6}, along the $y$-axis. The black dashed lines show the
Fermi energy.}\label{Figure7}
\end{figure}

\subsection{Resonances in 1-5 Ca dimer clusters}

The inclusion of the electric potential in our calculations has
crucial effects on the calcium-dimer-induced Dirac point resonant
states of the graphene monolayer that will be studied here. Before
presenting the main results for the LDOS, we should emphasize that
the influence of carbon $2s$, $2p_x$, and $2p_y$ orbitals on the
electronic states for the dimer-graphene system was examined. A
comparison of the resonances with and without the inclusion of
these orbitals showed that the contribution of C $2s$, $2p_x$, and
$2p_y$ states to the energy and intensity of the resonances is
very small and can be ignored in our calculation, indicating that
the C $2p_z$ orbitals make the main contribution in the emergence
of the resonance states. Therefore, in the Hamiltonians given in
Sec. II, only the C 2$p_z$ orbitals are retained in the LDOS
calculations. In addition, we shifted all of the electronic
spectra to set the graphene Dirac point at zero energy. In the
physical analysis of resonances, only the energy window close to
the graphene Dirac point has been shown. The LDOS for graphene in
the presence of Ca dimers at different lateral distances from the
center of each dimer-graphene cluster (see Fig. 1) is shown in
Figs. 4(a)-4(e). Note that in all clusters, the central dimer is
the first Ca dimer that is placed on the graphene monolayer and
the rest of the dimers surround it. All distances are measured
from the center of the first Ca dimer. Figure 4(a) shows that in
the case of one Ca dimer on the graphene layer, a weak resonance
and oscillation appears in the energy window of the plot; the
electronic states above the Fermi energy are affected the most by
the presence of the Ca dimer. These Ca-related features decrease
in strength with increasing distance from the Ca dimer (see the
LDOS in Fig. 4(a) at 8.48 nm). The LDOS also displays an
asymmetric behavior around the Dirac point energy that persists
even at large distances from the Ca dimer, partly due to the
inclusion of overlap matrix $S_{i,\alpha \beta}$ in our theory. In
this study, to find the Fermi energy for each dimer-graphene
system, the energy of highest occupied molecular orbital (HOMO)
was first determined for the central graphene ribbon with and
without Ca dimers. Then the difference between the two HOMO
energies was added to the Dirac point energy as our estimate of
the Fermi energy for the dimer-graphene system. This procedure was
repeated for all of the dimer-graphene clusters. The
electrochemical potential in the graphene leads was taken to be
the Dirac point energy renormalized to match the Fermi energy as
determined according to the above procedure.

In the case of the two-dimer cluster [Fig. 4(b)], the oscillation
in the electronic LDOS above the Dirac point energy is stronger
and has shifted downwards in energy. Moreover, the lowest energy
resonant state is now partly populated with electrons, signaling
that the atomic collapse is beginning to take place. This
situation resembles the onset of atomic collapse in the STM
experiments \cite{Wang}, but corresponds to the case of four Ca
dimers adsorbed on graphene in the experiment.

Possible reasons why the onset of atomic collapse occurred for a
larger number of Ca dimers in the cluster in the experiment of
Wang \textit{et al.} \cite{Wang} than in the present theory may
include the following: In the experiment\cite{Wang}, the graphene
layer was on a BN substrate, whereas in the present work, the
calculations were carried out for suspended graphene in vacuum
(with no substrate). Dielectric screening due to the substrate in
the experiment is expected to reduce the strength of the Coulomb
potential well induced in the graphene by the Ca dimers, thus
requiring a larger number of adsorbed dimers for the atomic
collapse to take place. The larger spatial separation between the
Ca dimers in the experiment than in the present theory is also
expected to result in a weaker Coulomb well in the experimental
system, with qualitatively similar consequences.

As the number of Ca dimers increases from two to five [Fig.
4(b)-4(e)], the lowest resonant state becomes completely occupied
and the number of quasi-bound states seen in the energy window in
Fig.4 increases. In fact, a series of resonances develops which
may be related to the Bohr-Sommerfeld quantization as was
discussed in Ref. \cite{Shytov1}; however, the resolution of the
STM experiment \cite{Wang} was only able to detect the lowest
state. As can be seen in Fig. 4, the energy of the lowest atomic
collapse state shifts below the graphene Dirac point as more
dimers are added to the group, while the Fermi level moves up in
energy. This means that additional charge transfer from Ca atoms
to graphene occurs in the process of adsorbing each additional Ca
dimer on the graphene monolayer.

In the experiment of Wang \textit{et al.} \cite{Wang}, a gaplike
feature in normalized differential conductance was reported which
comes from the effects of inelastic tunneling and the lifetime
broadening by electron-phonon and electron-electron interactions
\cite{Brar}. Since we are only interested in studying the
resonance features coming from the Ca dimers in suspended graphene
LDOS, these effects have not been included in our calculations.

To examine the spatial dependence of the resonant states, the
electronic LDOS were computed at different distances from the
center of each dimer-graphene cluster. The results are presented
in Fig. 4.  Although the intensities of the resonances in the LDOS
depend strongly on the distance from the center, the effect of
resonances on the electronic spectra is still discernible even
8.48 nm from the center (the black curves in Fig. 4). This
suggests that the resonant states are located mainly on the
graphene as distinct from the localized atomic orbitals on Ca
atoms.

To prove this point definitively, we removed the Ca atomic
orbitals from the our model tight-binding Hamiltonian but retained
the EPE values on the graphene that were obtained in our DFT
calculations for the systems consisting of the graphene and
adsorbed Ca dimers. In this way, only the effect of charge
transfer between the Ca atoms and graphene on the graphene
electric potentials was included in this test calculation, while
the role of Ca valence orbitals was ignored. Interestingly, the
results showed a negligible change in the electronic density of
states when compared with the results in Fig. 4, showing that the
presence of the resonance states is mainly due to the charge
transfer from Ca dimers to graphene and not due to the
hybridization between Ca $4s$ and $4p$ orbitals and graphene
$2p_z$ states.

\subsection{Resonances in a 6-dimer cluster and ring}

We explored the dependence of the Dirac point resonance features
on the spatial arrangement of the Ca dimers adsorbed on the
graphene by comparing the calculated LDOS for two different
structures, each composed of six Ca dimers on the graphene
monolayer. The dimer arrangements of the two structures considered
are shown in Figs. 5(a) and 5(b). If we add two more dimers to the
left and right side of the central dimer in Fig. 1(e), the
six-dimer cluster of Fig. 5(a) is obtained. On the other hand, if
the central dimer in Fig. 5(a) is shifted down along the diagonal,
the 6-dimer ring shown in Fig. 5(b) is formed. The calculated
local densities of states for the six-dimer cluster and the
6-dimer ring are shown in Figs. 6. The LDOS for the 6-dimer
structures is based on the EPE values whose averages are shown in
Fig. 3(c). The averaged EPE value for the six-dimer ring does not
change significantly as a function of $R$ for $R\leq 6${\AA}. This
is related to the absence of a central dimer in that structure and
is a completely different behavior than what is seen in Fig. 3(c)
for the other dimer-graphene clusters.

For both of the six-dimer structures the potential well in the
range of $R>7.5${\AA} is deeper than that for clusters composed of
1-5 Ca dimers, indicating more quasi-bound states for structures
of this type. Figures 6(a) and 6(b) clearly show two distinct
quasi-bound states near the Fermi level. Interestingly, the LDOS
in the two arrangements shows that the second atomic-collapse
state is becoming populated. We note that only a single atomic
collapse resonance was observed to be populated in the STM
experiment of Wang \textit{et al.} \cite{Wang}. However, Wang
\textit{et al.} \cite{Wang} did not report any studies of
six-dimer clusters. Although some of the the resonance features
are somewhat narrower for the six-dimer ring than that for the
six-dimer cluster in Fig. 6, it is evident that the energetic and
spatial characteristics of the three lowest LDOS resonances are
not affected greatly by the change in the arrangement of the
dimers. The most significant difference between the LDOS for the
two geometries is in the resonance energies above 0.2 eV at short
distances (2.21 nm) from the center of clusters.

These results suggest that the adsorption of more Ca dimers on the
graphene monolayer results in the transfer of more electric charge
from Ca atoms to the graphene monolayer and a deeper electrostatic
potential well in the graphene and that, accordingly, more atomic
collapse states will be populated.

\section{Conclusion}
In summary, we have studied theoretically the Coulomb bound states
and Dirac point resonances induced in a suspended graphene
monolayer due to the adsorption of different number of Ca dimers
on the graphene. We performed \textit{ab initio} DFT calculations
of the relaxed geometry for graphene with a single adsorbed Ca
dimer and also the Mulliken atomic charges and the atomic
electrostatic potentials for 1-6 Ca dimers on the suspended
graphene. The tight-binding extended H\"{u}ckel model, modified to
include the electrostatic potentials obtained with DFT, was then
used to compute the electronic structure of the system consisting
of the adsorbed Ca atom dimers and the carbon atoms of the
graphene monolayer. Our findings indicate that a charge transfer
from dimers to the graphene occurs and increases as the number of
dimers in the adsorbed cluster is increased from one to six.

Our calculated electronic structures reveal that the first
atomic-collapse state close to the Dirac point energy begins to
become populated with electrons when two Ca dimers are adsorbed on
the suspended graphene. By contrast in the recent STM experiment
\cite{Wang}, the first atomic-collapse state was observed to begin
to populate with electrons when four Ca dimers were present in the
Ca cluster adsorbed on the graphene. Possible explanations of this
difference may include dielectric screening of the Coulomb
potential well in the graphene due to the BN substrate underlying
the graphene monolayer in the experiment \cite{Wang} and a
shallower potential well due to the larger spacing between the Ca
dimers in the experimental system.

Our theory also predicts a second atomic-collapse state to begin
to populate with electrons when six Ca dimers are adsorbed on the
suspended graphene. No studies involving adsorbed Ca clusters with
six Ca dimers were reported by Wang {\em et al.}\cite{Wang}. The
experimental study\cite{Wang} also only found evidence of a single
atomic-collapse resonance becoming populated with electrons.
Therefore, experiments testing our prediction of a second
atomic-collapse state being populated with electrons for larger
numbers of Ca dimers adsorbed on suspended graphene would be of
interest. We note that suspended graphene monolayer
nanostructures, although without adsorbed Ca dimers, have already
been realized experimentally and their transport properties have
been measured.\cite{Tombros11}

The calculated spatial characteristics of the resonance features
confirm that only the charge transfer from Ca atoms to graphene is
responsible for the formation of the atomic-collapse resonances
and that the local Ca valence orbitals are not involved
significantly in the resonant states. Furthermore, it is found
that the spatial arrangement of Ca dimers within the adsorbed
dimer cluster does not strongly affect the formation of the
atomic-collapse state and the location of quasi-bound states that
are close in energy to the Dirac point, although the overall size
of the Ca dimer cluster may be important. These findings can be
used in future nanoscale devices where there is a need to confine
electrical charges to a small area.

This work was supported by NSERC, CIFAR, WestGrid, and Compute
Canada.

\end{document}